# An *operando* calorimeter for high temperature electrochemical cells

David Young,[a] Ariel Jackson,[a] David Fork,[b] Seid Sadat,[b] Daniel Rettenwander,[a] Jesse D. Benck,[a] Yet-Ming Chiang[a, *]

[a]Massachusetts Institute of Technology, Cambridge, MA 02139
[b]Google LLC., Mountain View, California, 94043
*Corresponding author. Email address: ychiang@mit.edu (Y. Chiang)

**Abstract**

*Operando* calorimetry has previously been utilized to study degradation, side reactions, and other electrochemical effects in electrochemical cells such as batteries at or near room temperature. Calorimetric data can provide important information on the lifetime and thermal properties of electrochemical cells and can be used in practical engineering applications such as thermal management. High temperature electrochemical cells such as solid oxide fuel cells or electrolyzers can also benefit from *operando* calorimetry, although to our knowledge no such unit has been developed commercially. Herein, we report an *operando* calorimeter capable of simultaneous calorimetry and electrochemistry at temperatures up to 1,000 °C and in both oxidizing and reducing atmospheres. The calorimeter is constructed by modifying a commercial apparatus originally designed to study high temperature electrochemical cells in various gas environments. We utilize a grey-box, nonlinear system identification model to analyze calorimetric data and achieve an electrochemical cell power sensitivity of 16.1±11.7 mW. This *operando* calorimeter provides the tools needed to study both the thermal and kinetic behavior of electrochemical cells at elevated temperatures.

Keywords: operando calorimetry, high temperature electrochemistry, solid oxide fuel cells, degradation, system identification

**1.     Introduction**

*Operando* calorimetry is widely used to study the properties of electrochemical cells, such as degradation of materials during electrochemical cycling and parasitic side reactions that lower electrochemical cell efficiency [1-3]. The decomposition rate of active materials through parasitic reactions can be used to predict the performance degradation and overall device lifetime. In addition, the development of thermal models for electrochemical cells as they degrade is useful for thermal management in practical settings [3].

Such studies have primarily focused on the characterization of batteries or other electrochemical cells that operate at or near room temperature. There is interest in extending *operando* calorimetry to higher temperatures and to either oxidizing or reducing atmospheres. In particular, fuel cells that operate in the 200-800 °C range under exposure to both highly reducing ($H_2$) and highly oxidizing ($O_2$) atmosphere are a promising technology [4, 5]. There is also developing interest in high temperature solid oxide electrolysis and $CO_2$ reduction to synthetic fuels [6, 7]. In all of these systems, solid electrolyte degradation is of utmost importance as the electrolyte affects both efficiency and life [5]. Many studies use thermal techniques such as differential scanning calorimetry to determine temperature stability windows and rates of degradation in the solid electrolyte [8-11]. However, these methods are performed *ex situ* rather than during operation of a fuel cell, which may change the conditions affecting electrolyte stability.

High temperature calorimeters are common but typically operate in the differential scanning calorimetry mode [12, 13], which is not amenable to simultaneous application of multiple reactive gas streams and electrochemistry. Therefore, it is not surprising that calorimeters with these capabilities have not previously been reported in the literature and are not readily available commercially. To build such a calorimeter is not straightforward. The materials required to construct the necessary calorimeter components, such as the calorimeter housing or vessel, the temperature probes, and the electrical connections, can be subject to attack by both $H_2$ and $O_2$, as well as by water vapor present. Furthermore, the different gas streams in a fuel cell must be well-separated to avert safety issues resulting from mixing at high temperature.

Herein, we report a calorimeter capable of *operando* measurements of high temperature electrochemical cells simultaneously exposed to both oxidizing and reducing gas environments. We modify a commercially available apparatus, the ProboStat$^{TM}$ (Norwegian Electro Ceramics AS), developed for the study of electrochemical cells such as fuel cells at high temperatures in various gas atmospheres, to perform *operando* calorimetry during operation. Such an apparatus enables simultaneous extraction of thermodynamic and kinetic parameters using both electrochemical and calorimetric methods. The calorimetric sensitivity for an operating solid state electrochemical cell is 16.1±11.7 mW. The details of this apparatus, its performance, and its potential uses are described below.

## 2. Experimental

*2.1. Design Objectives*

The calorimeter's temperature, atmosphere, and electrochemistry capabilities presented here are designed to support *operando* studies of high temperature solid electrolytes. The required performance parameters are presented in Table 1. The operating temperature range, from room temperature to 1,000 °C, permits use of the majority of solid electrolytes used in fuel cells. The electrical parameters allow for the application of electrochemistry and the use of temperature sensors for the calorimeter. With at least two temperature probes, each utilizing two connections, and a two electrode setup to probe the electrochemical cell or device, a minimum of 6 electrical connections are needed. If additional capabilities are needed, such as running three electrode measurements or four point probe tests, up to 11 electrical connections may be required. Designing the calorimeter to enable the use of two different gases simultaneously is necessary for testing fuel cells under realistic operating conditions. The samples used have a diameter of approximately 20 mm, a common size format in literature [8, 10].

A typical target for electrical power density in fuel cells is 500-2,000 mW/cm$^2$ [14]. Because currently fuel cells are approximately 50% efficient [15], a 500 mW/cm$^2$ thermal power density is assumed here, representing the low end of the range above. The goal of our calorimeter design is to capture at least 10% of that power. Therefore, the minimum sensitivity is 50 mW, assuming a minimum active electrode area of 1 cm$^2$. Furthermore, estimates from solid oxide electrolyte side reactions in the literature [10] show heat effects of order 100 mW/g. Assuming that samples contain 1 g of material, typical for the sample dimensions used here, heat effects are expected to be on the order of 100 mW for these reactions, which is measurable at the target sensitivity.

Table 1. Calorimeter design objectives for the study of high temperature solid electrolytes.

| Specification | Minimum Value | Maximum Value |
| --- | --- | --- |
| Temperature | 20 °C | 1,000 °C |
| Voltage | -20 V | 20 V |
| Current | 0 mA | 400 mA |
| Electrical connections | 6 | 11 |
| Number of gases | 1 | 2 |
| Sample diameter | 10 mm | 30 mm |
| Sensitivity | | < 50 mW |

*2.2. Calorimeter Design*

The calorimeter presented here modifies a commercially available high temperature electrochemistry apparatus, the ProboStat$^{TM}$ (Figs. 1 and A.1), which provides the necessary temperature range, electrical connections to power the fuel cell samples and temperature sensors, and two-gas capability. The apparatus tubing is made from either alumina for high temperatures (>600 °C) or quartz for medium temperatures (<600 °C), and the section exposed to elevated temperatures uses electrical connections made from platinum. Both materials remain inert and structurally stable in high temperature $H_2$ and $O_2$ ambient. Here the ProboStat$^{TM}$ is operated within a vertically mounted 18" tube furnace (Mellen) that can reach temperatures in excess of 1,000 °C.

Modifications to the ProboStat$^{TM}$ are made through the addition and relocation of temperature sensors. An S-type thermocouple that is provided with the ProboStat$^{TM}$ for monitoring the approximate sample temperature is relocated to the alumina support post below the sample and cemented (Cotronics Resbond 907GF) in place for better heat conduction. This thermocouple is used to measure the temperature proximal to the location of the sample ($T_{prox}$). A small, thin film resistance temperature detector (RTD, Omega F3142 or US Sensor PPG102A1) is attached to the sample itself, again using high temperature cement. In addition to the furnace thermocouple used by its PID temperature controller, a K-type thermocouple ($T_{fw}$) is attached to the wall of the furnace near the sample. $T_{fw}$ is used to acquire a more accurate temperature reading of the furnace temperature, picking up fluctuations in the wall temperature that are not recorded by the PID thermocouple. Finally the temperature of the room is monitored using a K-type thermocouple ($T_{room}$) located in the vicinity of the ProboStat$^{TM}$ base. All temperature data from thermocouples are recorded using an eight channel data acquisition module (Omega).

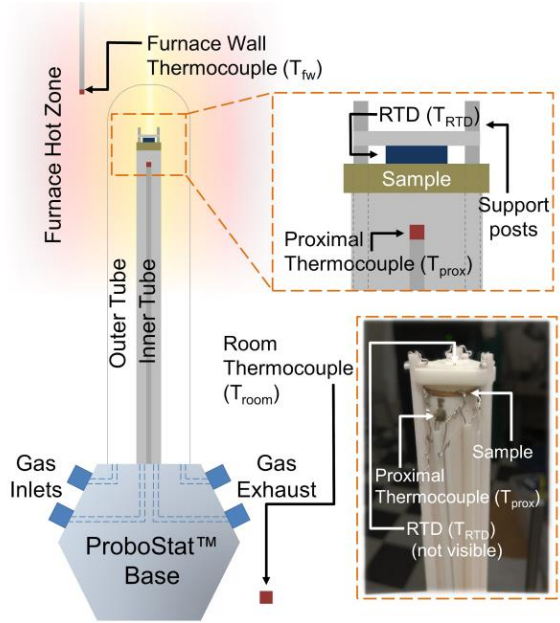

Fig 1. Schematic of a calorimeter, based off a ProboStat$^{TM}$ high temperature electrochemistry apparatus, for calorimetry of high temperature solid state proton conductors. A sample is mounted into the ProboStat$^{TM}$, which is placed in a vertical tube furnace. The apparatus is modified with several thermocouples to detect local furnace temperature ($T_{fw}$), the temperature proximal to the sample ($T_{prox}$), and the surrounding room temperature ($T_{room}$).

The electrical connection schematic is shown in Fig. 2a. The ProboStat$^{TM}$ is designed for a maximum of six electrical connections for power applications and up to six thermocouple connections. One pair of thermocouple connections is used for $T_{prox}$. All six power connections are used. Two of these electrical connections are used for operating the RTD, while four are connected to the sample. Two connections are made to opposite ends of the cathode to enable current flow across the cathode, as discussed in section 4.1, in order to simulate heat-generating reactions at the electrolyte-electrode interface. In addition, one connection is made to the anode, and another is made to the reference electrode of the sample. In certain cases, additional connections are needed for four point measurements or other types of electrical stimulation or monitoring. In those situations, voltage measurements can be made using the unused thermocouple connections, while reserving the other electrical connections for running current. All electrical sourcing and measurement is done on a potentiostat (Bio-Logic).

The calorimeter's fluidic infrastructure is designed for use of up to two gases with the option of running those gases dry or humidified (Fig. 2b). The sample, sealed to the inner tube of the ProboStat$^{TM}$, prevents leaking and mixing of the gases. A water bubbler is used to humidify gases at room temperature to about 3% (0.03 atm). The gases are constantly flowed through the ProboStat$^{TM}$ and vented out. Flow rates are typically set between 5-20 ml/min.

The calorimeter can support samples between 10-30 mm in diameter and between 0.5-10 mm in thickness, which corresponds to a sample mass of up to approximately 10 g. These samples are mounted into the Probostat$^{TM}$ with the appropriate electrical connections and can be sealed to the inner tube to allow for the simultaneous use of two gas atmospheres, each being exposed to one side of the sample. A more detailed explanation of sample design can be found below.

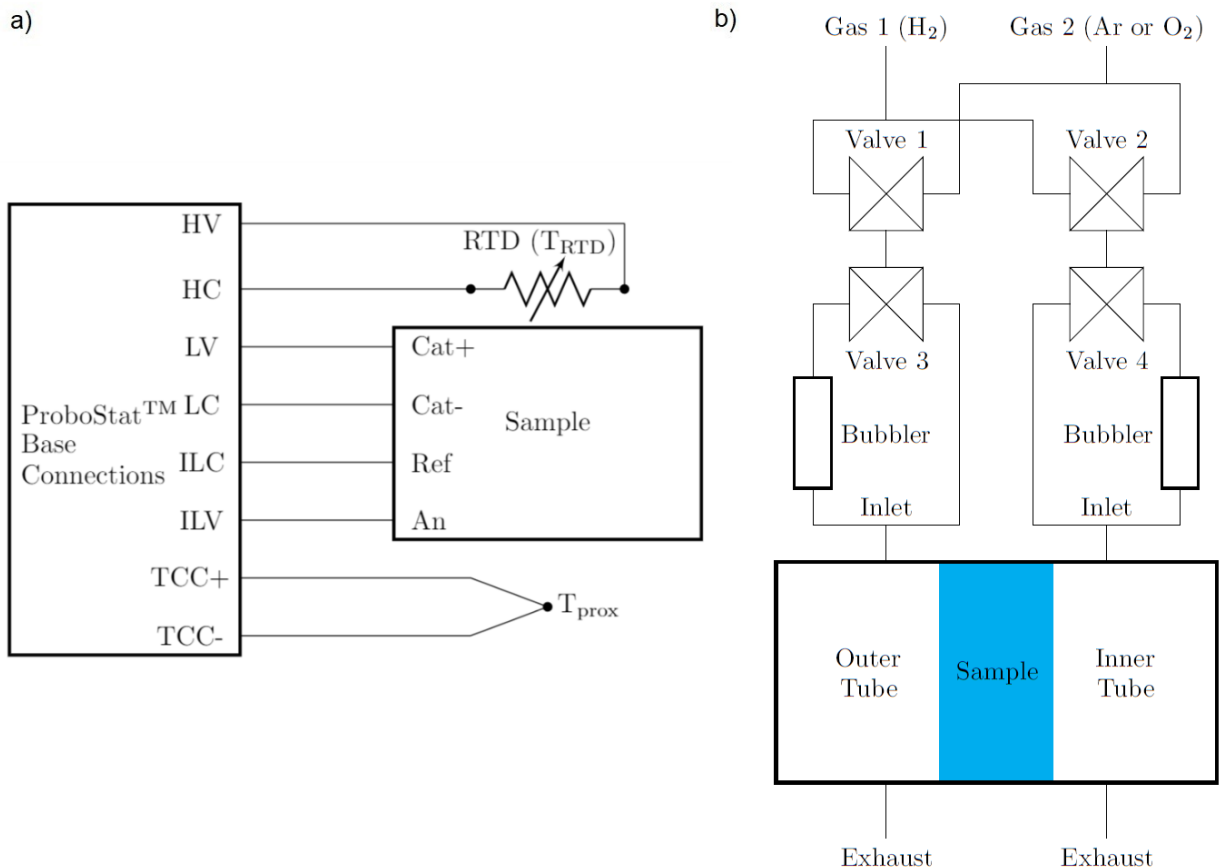

Fig. 2. (a) Electrical connection of the calorimeter. The ProboStat[TM] Base Connections correspond to connection ports labelled on the apparatus and are connected to various components of the setup. Two connections are made to a resistance temperature detector (RTD). Four connections are made to the sample: two to the cathode (Cat+ and Cat-), one to the anode (An), and one to the reference (Ref). Two connections are made to a thermocouple proximal to the sample ($T_{prox}$). (b) Fluidic diagram of the calorimeter. Two gas streams flow into the outer tube volume and the inner tube volume of the calorimeter. The sample can be sealed to separate the two gases if they are different. Two bubblers are placed in line to allow for humidified gas.

*2.3. Sample Design*

In order to test the high temperature electrochemistry performance of the calorimeter, samples were designed to mimic a working fuel cell. In this report, only $H_2$ is supplied to the sample in order to simplify operation for proof of concept of the calorimeter. However, in principle this same sample design could function as a fully operational fuel cell by supplying $H_2$ and $O_2$ to the anode and cathode respectively. In the case examined here, hydrogen gas is oxidized at the anode, supplying protons to the solid electrolyte which flow to the cathode where they are reduced to evolve hydrogen gas. No net molecules of hydrogen are produced or consumed in this operation, but ionic current does flow through the electrolyte, mimicking the electrochemical operation of a solid oxide fuel cell.

The sample is constructed using a 20 mm diameter and 1 mm thick $BaZr_{0.8}Ce_{0.1}Y_{0.1}O_3$ (BZCY) ceramic, proton conducting disk (CoorsTek Membrane Sciences AS) as a platform (Fig. 3). Thin film Pt and Pd electrodes are used as catalysts that promote the H evolution and oxidation reactions needed to supply protons to the BZCY disk. A cathode, anode, and reference between 20-200 nm are sputtered on top of a 3-4 nm sputtered Cr adhesion layer, using Kapton tape to mask off areas between the electrodes. To monitor the local temperature an RTD (Omega F3142 or US Sensor PPG102A1) is cemented on top of the cathode side of the sample, making sure that the RTD leads do not short to the cathode surface. The RTDs used in this report have a maximum operating temperature of 600 °C. However, RTD's with higher maximum operating temperatures are available (US Sensor PPG201B3 and PPG201E7) should one wish to study solid state electrochemical cells at higher temperatures.

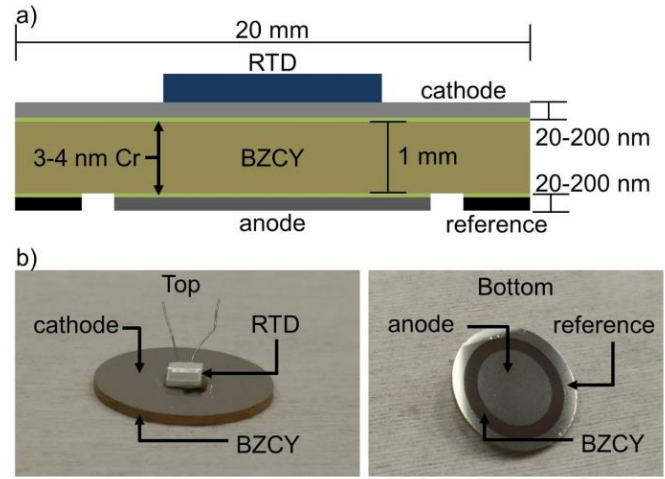

Fig. 3. (a) Schematic of sample with mounted resistance temperature detector (RTD). The sample contains a 20-200 nm Pd or Pt thin film cathode separated from a 20-200 nm Pd or Pt thin film anode and a ring-shaped reference by a 20 mm diameter and 1 mm thick BaZr$_{0.8}$Ce$_{0.1}$Y$_{0.1}$O$_3$ (BZCY) ceramic electrolyte. 3-4 nm of Cr is used as an adhesion layer between the metal electrodes and ceramic electrolyte. (b) Photos of the top and bottom of the sample.

## 3. Theory

*3.1. Inferring Heat Flow with a Nonlinear Lumped-Parameter Model*

A nonlinear lumped-parameter model is used to describe the behavior of heat transfer within the calorimeter, similar to the method described by MacLeod et al. [16]. The dynamics of this system are approximated using a grey-box approach. This entails developing an appropriate equivalent circuit model to describe the heat flow pathways through the calorimeter and then empirically deriving the parameters of the model. Temperature changes at various temperature probes resulting from known amounts of input power during model calibration determine the model parameter values that define the relationship between power and temperature in the model. This calibrated model can then be used to predict unknown amounts of evolved output power later during experiments knowing only the input powers and temperatures of those temperature probes. Any mismatch between the input and output powers are attributed to side reactions or degradation, which can be analyzed along with electrochemical data to gain insight on the performance of the solid electrolyte over time.

*3.2. Nonlinear equivalent circuit model of the calorimeter*

A one state, nonlinear equivalent circuit diagram of our calorimeter is shown in Fig. 4a. Numerous model candidates were analyzed, including two and three state models, but this one state model with two nodes and three measured grounds was determined to be the model that provides the best calorimeter sensitivity.

In this equivalent circuit model, the resistors represent heat conductances between different nodes. A capacitor represents a heat capacitance, while grounds represent thermal sources and sinks. Power inputs are modelled as current sources. The model has one measured node, $T_{RTD}$, and one unmeasured node, $T_{sample}$, although only one heat capacitance is needed to approximate both. Other than the absolute ground used to reference power inputs and heat capacitance, there are three grounds in this model: one source and two sinks.

Fig. 4b shows a schematic representation of the sample, supports, and temperature sensors along with the heat flow pathways in the system. This diagram illustrates the physical origins of the heat flows within the system that are represented in the equivalent circuit diagram (Fig. 4a). Heat is generated in the furnace, represented by $T_{fw}$, and flows to the RTD node, which has a temperature $T_{RTD}$. Heat then continues to flow through the sample node, an unmeasured node that approximates the sample itself, before exiting out either through an intermediate ground $T_{prox}$ or the room temperature ground $T_{room}$. The need for two sinks stems from the inability of $T_{prox}$ to capture all of the heat flowing out from the sample node due to other heat flow pathways out to the surrounding environment.

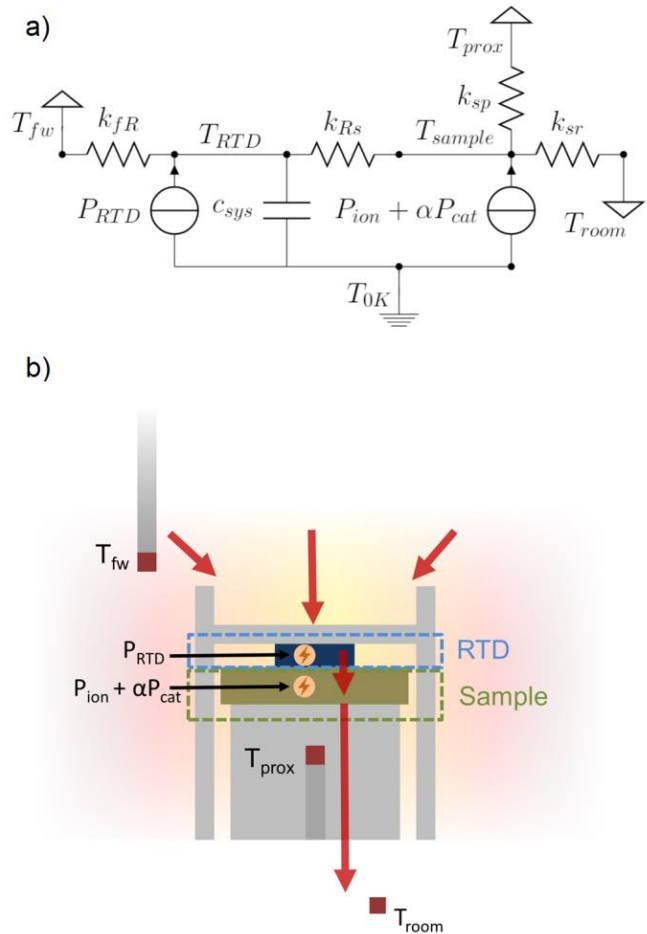

Fig. 4. (a) Equivalent circuit diagram of the calorimeter. The model consists of two nodes represented by $T_{RTD}$ and $T_{sample}$ but just one state capacitor $c_{sys}$ for the system. There are three signal grounds represented by the furnace temperature $T_{fw}$, the proximal thermocouple temperature $T_{prox}$, and room temperature $T_{room}$. One additional absolute ground $T_{0K}$ is used as a reference for heat capacitance. Powers from three different sources located at or near the sample are presented by $P_{RTD}$, $P_{ion}$, and $P_{cat}$. (b) Physical model of heat flow within the calorimeter. Red

arrows indicate the direction of heat flow starting from the furnace hot zone to the outside environment. Physical approximations of the nodes corresponding to $T_{RTD}$ and $T_{sample}$, as well as $P_{RTD}$, $P_{ion}$, and $P_{cat}$ labelled in (a) are also shown.

The model includes three power sources that correspond to the different power inputs to the sample during either calibration or testing. $P_{RTD}$ is a baseline power that is emitted by the RTD when in operation. Typically for a 1 k$\Omega$ RTD a sense current of 1 mA is used, generating between 1-5 mW of power, depending on the temperature of the system. $P_{ion}$ is the power generated from operating the solid electrolyte. It is caused by both resistive heating of the solid electrolyte and reaction overpotential losses. Lastly, $P_{cat}$ is used to study surface reactions or other surface phenomena related to the solid electrolyte. This power is approximated during calibration of the model by running current laterally across the cathode of the sample. Two nodes were modelled to appropriately capture system dynamics. The time constant for heat detection by the RTD for a heat source originating within the RTD ($P_{RTD}$) is smaller than the time constant for heat detection when the source is nearby, but not within, the RTD itself ($P_{ion}$ and $P_{cat}$).

The dynamics of the system described by the equivalent circuit diagram can be distilled into the following system of equations:

$$\frac{dT_{RTD}}{dt} = \frac{1}{c_{sys}}(P_{RTD} + k_{fR}(T_{fw} - T_{RTD}) - k_{Rs}(T_{RTD} - T_{sample})), \quad (1)$$

$$P_{ion} + \alpha P_{cat} + k_{Rs}(T_{sample} - T_{RTD}) + k_{sp}(T_{sample} - T_{prox}) + k_{sr}(T_{sample} - T_{room}) = 0, \quad (2)$$

where

$$c_{sys} = c_{sys,0} + c_{sys,1}T_{RTD} + c_{sys,2}T_{RTD}^2, \quad (3)$$

$$k_{fR} = k_{fR,0} + k_{fR,1}T_{RTD} + k_{fR,2}T_{RTD}^2, \quad (4)$$

$$k_{Rs} = k_{Rs,0} + k_{Rs,1}T_{RTD} + k_{Rs,2}T_{RTD}^2, \tag{5}$$

$$k_{sp} = k_{sp,0} + k_{sp,1}T_{sample} + k_{sp,2}T_{sample}^2, \tag{6}$$

$$k_{sr} = k_{sr,0} + k_{sr,1}T_{sample} + k_{sr,2}T_{sample}^2, \tag{7}$$

$$\alpha = \alpha_0 + \alpha_1 T_{sample} + \alpha_2 T_{sample}^2. \tag{8}$$

Although Eq. 1 and 2 could be combined into one equation, thereby eliminating $T_{sample}$, they are split out to better represent the physical model. Each element in the model is split out to second order nonlinear parameters, which aggregates to 18 possible parameters to be fitted. For operation within a limited temperature range, the model typically does not need to be fit with second order nonlinear parameters, so just zeroth and sometimes first order parameters are used. In addition to fitting capacitances ($c_x$) and conductances ($k_{xx}$), a scale factor ($\alpha$) is fitted to adjust $P_{cat}$ based on the ratio of $P_{cat}$:$P_{ion}$ detected by the temperature probes in the calorimeter.

## 4. Results and Discussion

### 4.1. Model calibration and parameter estimation

Prior to running test experiments, a calibration is performed on the calorimeter using either a dummy sample or the test sample itself. The calibration is used to estimate the model parameters by applying temperature and power inputs that span in magnitude and frequency beyond the limits of those imposed during the actual experiment. A calibration needs to be performed before every experiment due to parameter variation from one run to the next, which can be attributed to small shifts in the position of the Probostat™ relative to the furnace, the sample, or the support fixtures used to mount the sample.

An example of a calibration is shown in Fig. 5. In this case, both input powers (Fig. 5a) and temperature (Fig. 5b) are varied over the course of the calibration. Three input powers $P_{RTD}$, $P_{cat}$, and $P_{ion}$ represent three possible sources of heat generated during an experiment. The temperatures $T_{fw}$, $T_{prox}$, and $T_{room}$ (not shown) are the grounds in the model from Fig. 4a. $T_{RTD,meas}$ is the measured temperature at the RTD node, $T_{RTD}$ in the model from Fig. 4a. As mentioned earlier, $T_{sample}$ is not measured and is instead estimated during the model fitting. The input data from Fig. 5a and 5b are analyzed with a MATLAB script utilizing the MATLAB System Identification Toolbox™ *nlgreyest* procedure similar to one outlined by MacLeod et. al. [16, 17]. The script takes the model described by Eq. 1-8, as well as the input power and ground temperatures, and estimates the vector of model parameters θ to create a modelled version of $T_{RTD}$, $T_{RTD,mod}$. A search for fitted model parameters is initiated and iterated over in order to minimize a cost function

$$V(\theta) = \frac{1}{t_{max}} \sum_{t=0}^{t_{max}} (\Delta T_{meas}(t) - \Delta T_{mod}(t,\theta))^2, \tag{9}$$

$$\Delta T_{meas}(t) = T_{RTD,meas}(t) - T_{prox}(t), \tag{10}$$

$$\Delta T_{mod}(t,\theta) = T_{RTD,mod}(t,\theta) - T_{prox}(t), \tag{11}$$

where t is the discretized time of each datapoint in the dataset ranging from 0 to $t_{max}$. A differential measurement between the measured or modelled temperature at the RTD node and a nearby temperature ground allows for fits to be made on smaller temperature ranges and cancels out the majority of noise coming from the furnace, both of which improve the goodness of fit. Typically only linear (zeroth order) terms for each element in the model are fitted at first to scan for a global minimum in V(θ) with lower computational effort. Upon approaching a perceived cost minimum, first order, nonlinear terms for each circuit element can be allowed to vary as well. Second order, nonlinear terms are rarely used unless the experiment spans a temperature range of over several hundred °C.

After a suitable search and refinement of the model parameters θ that either minimizes V(θ) below a certain cost tolerance or reaches a maximum number of search iterations, θ is said to be equal to $\theta_{minimized}$. While the goal is to reach a global minimum for the function V(θ), it is likely that $\theta_{minimized}$ represents a local minimum that is near the global minimum but may not be the absolute global minimum attainable within the given model structure. Upon reaching V($\theta_{minimized}$), $\Delta T_{meas}$ is compared to $\Delta T_{mod}$ (Fig. 5c). An example of parameters for a $\theta_{minimized}$ are shown in Table A.1 of the supplementary material. A goodness of fit is established by taking the normalized root mean square error (NRMSE) between the $\Delta T_{meas}$ and $\Delta T_{mod}$ datasets. While the NRMSE is a good screening metric, one should be cautious about ascribing too much value to it since it is easily affected by the dynamic range of the temperature differential dataset. A dataset with a large dynamic range in temperature differential can lead to a better NRMSE score than an equivalent calorimeter with a lower dynamic range in the calibration dataset. A better quantification for the quality of the calorimeter-model system is the power sensitivity, which is described in the following section. Nonetheless, it is a useful feedback tool during parameter honing with a calibration dataset. Typically an acceptable result is above 60%, with the best calibrations exceeding 75% NRMSE. In the case shown in Fig. 5c, the NRMSE is 79%. Poor sets of model parameters are easily identified with the NRMSE metric. NRMSE ranges from negative infinity to 100%, so negative NRMSE values (such as -75%) are not uncommon in the early iterations of parameter searching, which indicates a fit worse than a line averaging the dataset.

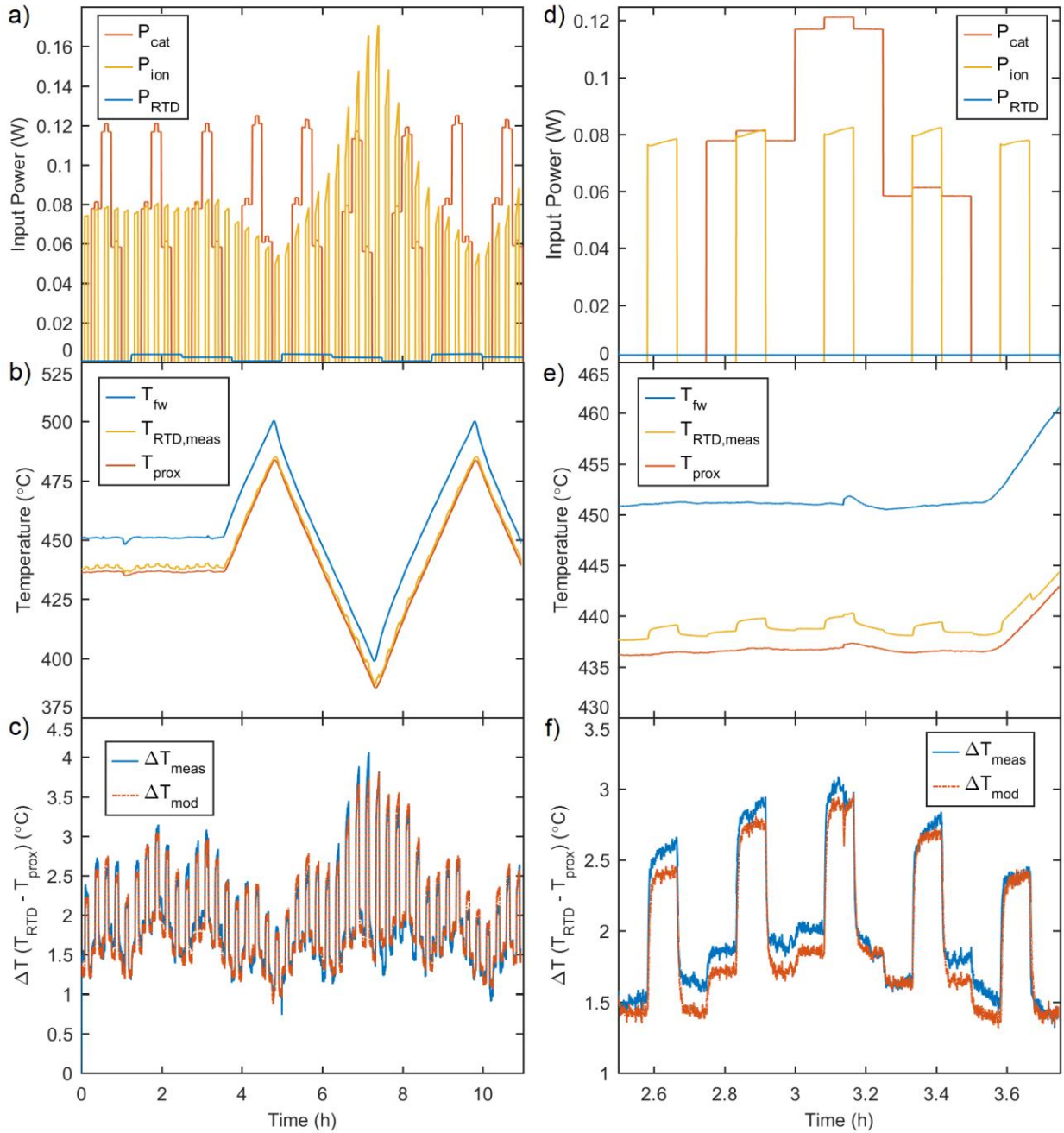

Fig. 5. An example calibration run on a Pd thin film coated BZCY sample at elevated temperatures in humidified $H_2$ gas for determining model parameters. (a) Calibration input powers from three input sources, (b) temperature outputs, and (c) a fit between the modelled temperature and the measured one. The normalized root mean square error (NRMSE) of the fit is 79%. (d)-(f) are excerpts of (a)-(c) magnified over a shorter time span to demonstrate finer details of the signals.

*4.2. Determination of calorimeter sensitivity*

To determine the performance of the fitted model on other datasets, a second experiment, called a prediction, is run and analyzed. A set of input power signals utilizing different waveforms and magnitudes are used to verify that the model is applicable to a wide variety of datasets, not just the one used for calibration. The input powers from part of a prediction run are highlighted in Fig. 6a, along with temperature in Fig. 6b. This time instead of trying to fit $\theta$, the parameters are fixed to $\theta_{minimized}$ from the previous calibration. $T_{RTD,meas}$ is used to in conjunction with the measured temperature grounds and the modelled parameters to recover a modelled version of the input power. This modelled power can be compared with the original input power to see how well the model reconstructs the power data (Fig. 6c). Residual power is calculated from the difference between these two powers (Fig. 6d). The standard deviation of the residual power indicates the calorimeter's lower limit of detection. However, an adjustment needs to be made to account for the fact that power is measured in the RTD node in the model and the powers of interest ($P_{ion}$ and $P_{cat}$) originate in the sample node. Only a fraction of the sample node's power flows into the RTD node, so the power signal and also noise in the power is dampened. The standard deviation is multiplied by an adjustment factor from a node-to-node heat flow analysis described in Appendix B of the supplementary material. Doing so approximates the noise level in the sample node that is relevant to $P_{ion}$ and $P_{cat}$. In the prediction example presented here, the standard deviation is 1.7 mW and the adjustment factor is 2.6, giving an adjusted standard deviation of 4.4 mW.

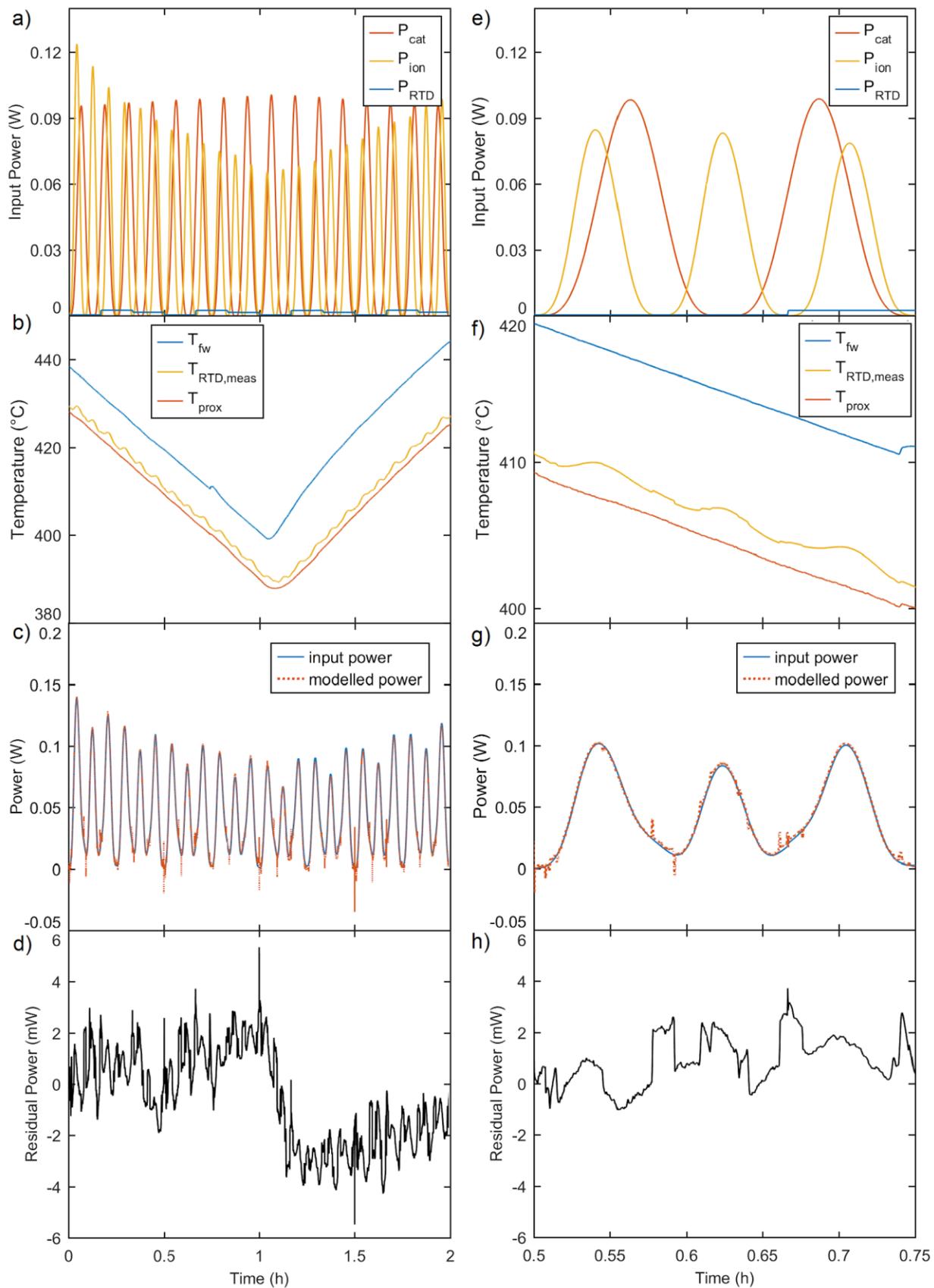

Fig. 6. An example prediction run on a Pd thin film coated BZCY sample at elevated temperatures in humidified $H_2$ gas, used to verify calibration parameters and determine calorimeter sensitivity. (a) Prediction input powers from three input sources utilizing different waveforms, (b) temperature outputs, (c) a fit between the modelled power and measured one, and (d) the residual power taken from the difference between the input and modelled powers. The standard deviation of the residual power (1.7 mW in this case) is used to determine the calorimeter sensitivity. (e)-(h) are excerpts of (a)-(d) magnified over a shorter time span to demonstrate finer details of the signals.

After using this calibration procedure, the calorimeter can be used to test solid electrolyte samples for any heat released from degradation or side reactions during application of electrochemistry. Power signals observed during experiments beyond two times the adjusted standard deviation found in the prediction step (8.8 mW in the above case) are considered to be heat generated from reactions unexplained by the input powers into the system with 95% confidence. This metric is defined as the calorimeter sensitivity. These excess power signals also need to be adjusted by multiplying by the adjustment factor to get the true power coming from the sample.

A set of seven calibration and predictions steps were performed over different samples to assess the average performance of the calorimeter (Fig. 7). The calorimeter was found to have an average sensitivity of 16.1±11.7 mW, which meets the design objective of 50 mW. The performance of the calorimeter varied between samples, which could be due to some differences in sample preparation and mounting into the calorimeter. Average power residuals during control testing of samples were found to be well below the sensitivity for all samples, demonstrating the ability of the model with parameters $\theta_{minimized}$ to properly account for all measured temperature fluctuations based on the known input powers. These power residuals were taken by averaging all of the power residuals during a prediction run and represent bias from noise or inaccuracies of the model in predicting the expected power.

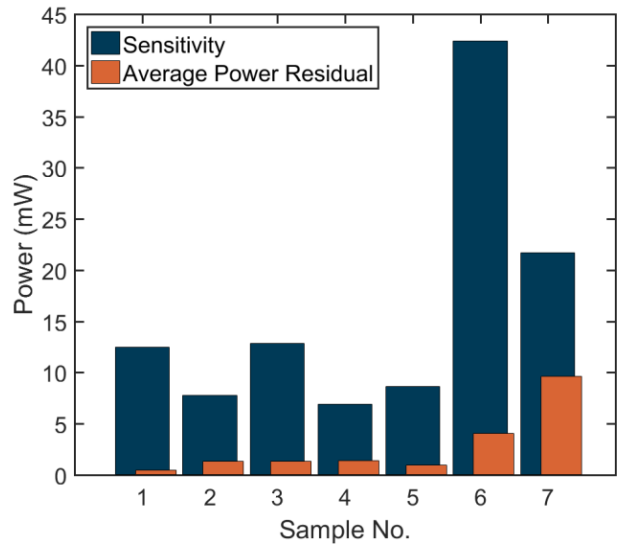

Fig. 7. Calorimeter sensitivity and average power residuals of seven control runs. Calorimeter sensitivity is defined as the adjusted residual power level above which heat unexplained by known input powers is observed. The average calorimeter sensitivity is determined to be 16.1±11.7 mW. The average power residuals are the adjusted power residuals seen over the course of a prediction experiment. Deviations from zero are attributed to noise in the system or inaccuracies of the model. In all seven control runs, the average power residual is well within the sensitivity for that run, as is expected for a control run, demonstrating that the model is able properly account for all measured temperature fluctuations as a function of the input powers.

*4.3. Extending Measurement Capabilities*

The calorimeter proposed here represents a step towards performing calorimetry under the challenging conditions of simultaneous high temperature, multiple gas environments, and application of electrochemistry. The calorimeter is sensitive to around 10-20 mW of power and is convenient to construct from a commercial apparatus that is well known to the high temperature fuel cell community. In comparison, top of the line commercial calorimeters can detect on the order of 10-100 µW, but these are units that are designed to optimize sensitivity rather than flexibility of operating conditions and can typically operate up to only ~100 °C. Admittedly, further improvements can be made to our calorimeter, especially to detect smaller heat signatures from side reactions and electrolyte degradation that may be present below the detection limit.

Certain modifications can be made to improve the resolution of this apparatus. The biggest shortcoming of this calorimeter, as designed, is that the RTD mounted onto the sample is not capturing enough of the heat being generated at the sample. To increase signal temperatures at the sample's RTD, a decrease in conduction of heat from the sample to both the furnace wall and the room temperature ground is desired, leaving more heat to enter the RTD for a given power input. A thinner alumina tube and support pieces in contact with the sample would reduce the main pathway for conductance from the sample out to heat sinks. Similarly, thinner contact wires for electrochemistry will also reduce conductance. Alternatively, introducing a thermal insulator (such as an aerogel) between the sample and the support tube would help to thermally isolate the sample from the rest of the apparatus.

In addition to increasing the sample RTD sensitivity to input power, substantial gains could be had by sensing more of the heat flowing out of the sample at the proximal temperature sensor.

Currently, the single proximal thermocouple $T_{prox}$ only covers part of the alumina tube extending from the sample down the Probostat and does not capture inhomogeneous heat signatures emitted from the sample. By placing more thermocouples or using a highly heat conductive material to average out temperatures circumferentially around the tube, more of the heat flow out of the sample can be captured.

Smaller improvements may be possible by improving furnace stability and stabilizing the room temperature to reduce thermal noise in the system. Additionally, better electrical shielding in the ProboStat$^{TM}$ would reduce electrical noise on the sample and temperature sensors.

Finally, run-to-run variations can be minimized by offering less spatial freedom for mounting the sample. By designating exact locations for mounting fixtures to be placed, as well as for where the ProboStat$^{TM}$ sits in relation to the furnace, model parameters should vary less from one experiment to the next or could even be reused across experiments.

## 5. Conclusions

A calorimeter has been presented here to study *operando* high temperature solid state electrochemistry, namely solid electrolytes for fuel cells and other similar applications. It can operate between room temperature and 1,000 °C and can support up to 12 electrical and thermocouple connections, as well as expose samples to two different gas environments. The calorimeter has a sensitivity of 16.1±11.7 mW and can be constructed from a commercial apparatus and temperature sensors.

Additionally, a heat transfer equivalent circuit model and grey-box system identification technique has been described that allows for the estimation of system parameters and reproducible analysis of experimental data. This technique gives flexibility to the modification of both sample and apparatus design. In most circumstances, only model parameters are changed, and, in the worst case, the model is modified to include additional dynamics to the system, though the overall procedure for data analysis remains the same.

This calorimetry system provides the means to directly study the thermal characteristics and dynamics of solid electrolytes *operando*. These findings, when coupled with electrochemical data, can provide insights into electrochemical cell performance such as efficiency or power degradation, device life, and cell heating.

**Acknowledgements**

We thank Matt Trevithick and Dr. Ross Koningstein for helpful discussions. Financial support was provided by Google LLC. David Young was supported by the National Science Foundation Graduate Research Fellowship under Grant No. 1122374.

**Supplementary Material**

**Appendix A: Calorimeter and its equivalent circuit model parameters**

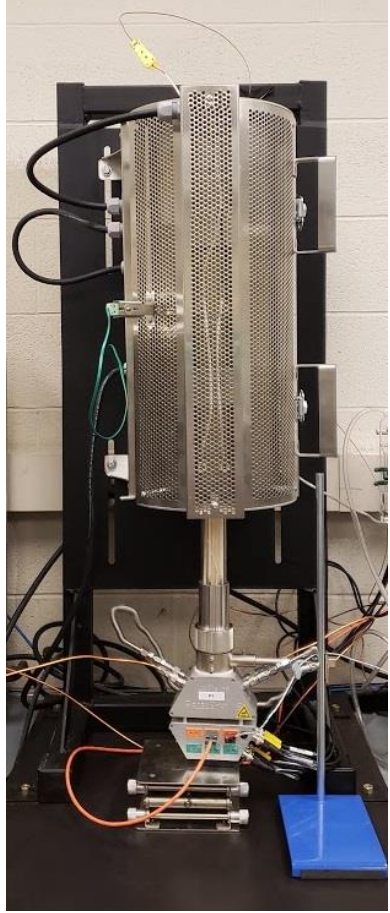

Fig. A.1. The ProboStat$^{TM}$ calorimeter setup inside a vertically mounted furnace and accompanying gas and electrical hookups.

Table A.1. Fitted parameters* obtained from a calibration utilizing MATLAB's *nlgreyest* function to fit calibration input powers and measured temperatures to a calorimeter equivalent circuit model.

| Parameter | Calibration Value | Uncertainty |
|---|---|---|
| $c_{sys,0}$ | 2.19E-1 J/K | 6.2% |
| $k_{Rs,0}$ | 1.31E-2 W/K | 4.9% |
| $k_{fR,0}$ | 4.84E-3 W/K | 4.6% |
| $k_{fR,1}$ | -5.42E-6 W/K$^2$ | 5.1% |
| $k_{sp,0}$ | 7.11E-2 W/K | 1.0% |

| α | 3.10E-1 | 2.9% |

*Parameters not included here were not used (i.e. value of zero) in this model fitting.

**Appendix B: Determination of the adjustment factor used to derive calorimeter sensitivity**

During calibration and prediction steps that determine calorimeter model parameters and calorimeter sensitivity, a direct analysis is made on the prediction power residuals to determine their standard deviation. This process is valid, but the resulting magnitude of the residual standard deviation assumes that 1 mW of any input power is equivalent to 1 mW of output power detected by the model. However, in reality, the sensitivity of the model to the ionic power is not the same as its sensitivity to the RTD power because the ionic power is introduced to the model at a different node. By the time the power from the sample node reaches the RTD node, the measured node, it has been dampened by the conductances in the model. In other words, not all power from the sample node flows into the RTD node, so an adjustment factor needs to be created to ascertain the true power sensitivity to the power generated in the sample node. These powers happen to be the ones of interest.

To assess this sensitivity, a calibration or prediction run is reanalyzed, keeping the same fitted parameters and input powers as before, except that the input ionic power is constrained to be 0 for the analysis. The result is shown in Fig. B.1. Because the model does not pick up any ionic power, almost all of the power residuals that appear are due to the ionic power. By measuring the peaks of the residuals and corresponding those to the actual input power, a relative power sensitivity can be obtained. For example, a peak of 25 mW in Fig. B.1b corresponds to a known input power of 170 mW in Fig. B.1a, which suggests that only 15% of the ionic power is collected by the RTD at the RTD node of the model (called the collection factor). Because the sensitivity is determined by the standard deviation of the power residuals, an adjustment factor is calculated using the following equation:

$$adjustment\ factor = \frac{1}{\sqrt{collection\ factor}} \qquad (B.1)$$

In this case, the adjustment factor is 2.6. Therefore, a 1.7 mW sensitivity displayed in the model is actually a 4.4 mW sensitivity to the ionic power.

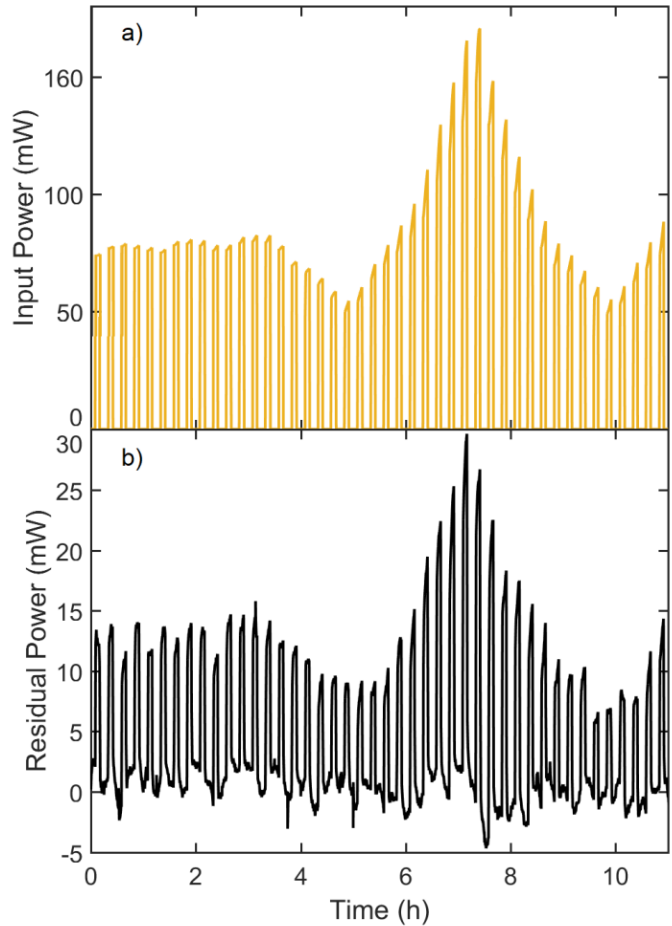

Fig. B.1. Comparison between (a) the ionic input power and (b) the power residuals observed from running a model fit and setting that ionic input power to zero, while keeping all other parameters, measured temperatures, and input powers the same. From this comparison, a collection factor of 15% is calculated, suggesting that only 15% of the ionic power is detected by the model's temperature measurements.

**Appendix C: Verification of Electrochemical Operation at Elevated Temperature**

Several preliminary studies were performed on testing the operation of BZCY samples to demonstrate various electrochemical capabilities of this system for studying fuel cells. For example, the conductivity of one of these samples exposed to humidified 4% $H_2$ in Ar at different temperatures is plotted in Fig. C.1 and voltage response to applied current at a range of temperatures is shown in Fig. C.2, demonstrating electrochemistry capabilities of the apparatus up to at least 800 °C.

From these, and other, initial performance verification studies, the voltage noise resolution is determined to be within 2 mV. Temperature variation on the furnace wall is less than 0.2 °C (standard deviation) at a fixed PID temperature controller setpoint of 400 °C over ~11 hours. However, an offset in the temperature control of the furnace is apparent. The temperature measured by $T_{fw}$ can be off by as much as +20-40 °C when compared to the furnace setpoint temperature when operating at high temperatures. Therefore, $T_{fw}$ is used as the actual temperature of the furnace for the purposes of performing calorimetry.

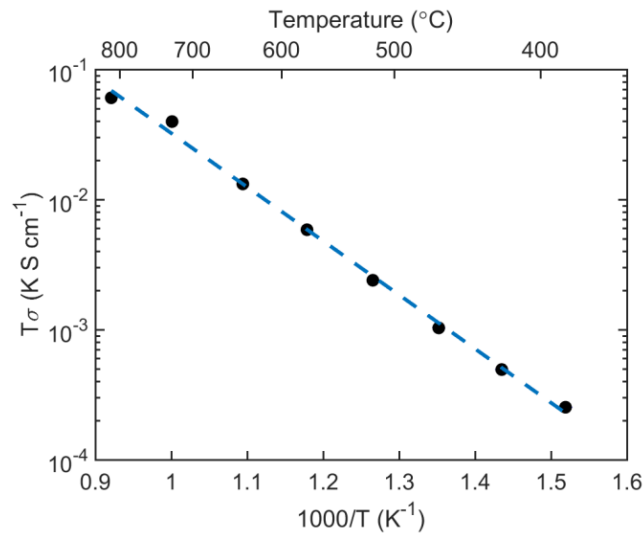

Fig. C.1. Demonstration of high temperature electrochemistry in the calorimeter. Conductivity of a BaZr$_{0.8}$Ce$_{0.1}$Y$_{0.1}$O$_3$ (BZCY) solid electrolyte with sputtered thin film Pt electrodes vs. temperature in humidified 4% H$_2$ in Ar. The activation energy of this BZCY sample is found to be 79 kJ/mol.

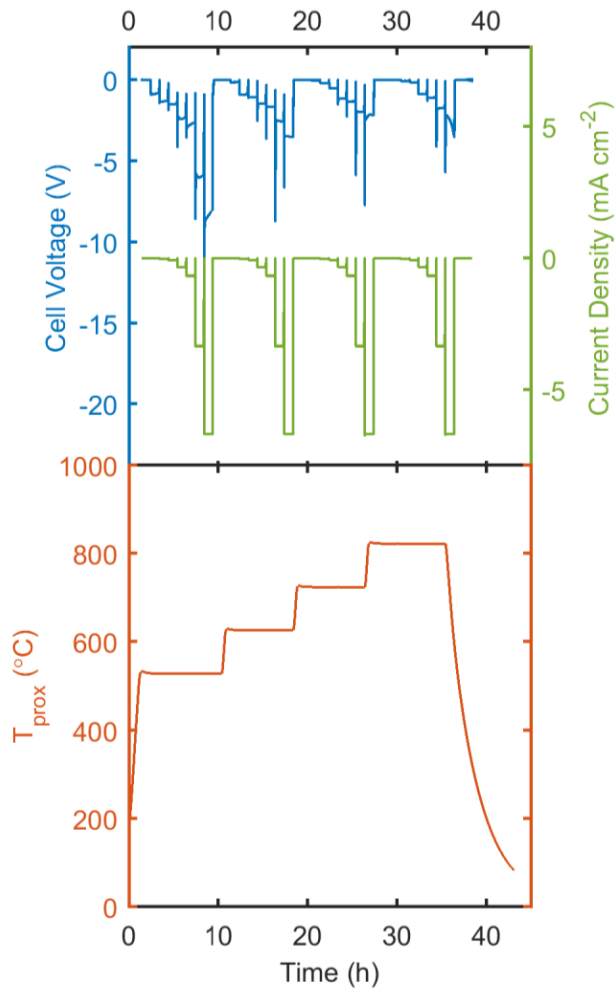

Fig. C.2. Demonstration of sustained high temperature electrochemistry in the calorimeter. Current is applied at temperatures between 500-800 °C in humidified 4% H$_2$. The observed voltage is primarily due to ohmic resistance in the BZCY electrolyte, explaining why the magnitude of voltage decreases with increasing temperature.

**References**

[1] L.E. Downie, J.R. Dahn, Determination of the Voltage Dependence of Parasitic Heat Flow in Lithium Ion Cells Using Isothermal Microcalorimetry, J. Electrochem. Soc. 161 (12) (2014) A1782-A1787.

[2] S.L. Glazier, L.E. Downie, J. Xia, A.J. Louli, J.R. Dahn, Effects of Fluorinated Carbonate Solvent Blends on High Voltage Parasitic Reactions in Lithium Ion Cells Using OCV Isothermal Microcalorimetry, J. Electrochem. Soc. 163 (10) (2016) A2131-A2138.

[3] Y. Saito, M. Shikano, H.Kobayashi, Heat generation behavior during charging and discharging of lithium-ion batteries after long-time storage, J. Power Sources 244 (2013) 294-299.

[4] N. Mahato, A. Banjeree, A. Gupta, S. Omar, K. Balani, Progress in material selection for solid oxide fuel cell technology: A review, Prog. Mater. Sci. 72 (2015) 141-337.

[5] B.C.H. Steele, A. Heinzel, Materials for fuel-cell technologies, Nature, 414 (2001) 345-352.

[6] M.A. Laguna-Bercero, Recent advances in high temperature electrolysis using solid oxide fuel cells: A review, J. Power Sources 203, (2012) 4-16.

[7] S.D. Ebbesen, M. Mogensen, Electrolysis of carbon dioxide in Solid Oxide Electrolysis Cells, J. Power Sources 193 (1) (2009) 349-358.

[8] N. Yan, Y. Zeng, B. Shalchi, W. Wang, T. Gao, G. Rothenberg, J. Luo, Discover and Understanding of the Ambient-Condition Degradation of Doped Barium Cerate Proton-Conducting Perovskite Oxide in Solid Oxide Fuel Cells, J. Electrochem. Soc. 162 (14) (2015) F1408-F1414.

[9] D.A. Boysen, T. Uda, C.R.I. Chisholm, S.M. Haile, High-Performance Solid Acid Fuel Cells Through Humidity Stabilization, Science, 303 (5654) (2004) 68-70.

[10] C. Chervin, R. S. Glass, S. M. Kauzlarich, Chemical degradation of $La_{1-x}Sr_xMnO_3/Y_2O_3$-stabilized $ZrO_2$ composite cathodes in the presence of current collector pastes, Solid State Ionics 176 (1-2) (2005) 17-23.

[11] F. Lufrano, I. Gatto, P. Staiti, V. Antonucci, E. Passalacqua, Sulfonated polysulfone ionomer membranes for fuel cells, Solid State Ionics 145 (1-4) (2001) 47-51.

[12] C. Toffolon-Masclet, T. Guilbert, J. C. Barchet, Study of secondary intermetallic phase precipitation/dissolution in Zr alloys by high temperature–high sensitivity calorimetry, J. Nucl. Mater. 372 (2-3) (2008) 367-378.

[13] Y. Moriya, A. Navrotsky, High-temperature calorimetry of zirconia: Heat capacity and thermodynamics of the monoclinic–tetragonal phase transition, J. Chem. Thermodynamics 38 (3) (2006) 211-223.

[14] E.D. Wachsman, K.T. Lee, Lowering the Temperature of Solid Oxide Fuel Cells, Science 334 (6058) (2011) 935-939.

[15] O.Z. Sharaf, M.F. Orhan, An overview of fuel cell technology: Fundamentals and applications, Renew. Sust. Energ. Rev. 32 (2014) 810-853.

[16] B.P. MacLeod, P.A. Schauer, K. Hu, B. Lam, D.K. Fork, C.P. Berlinguette, High-temperature high-pressure calorimeter for studying gram-scale heterogeneous chemical reactions, Rev. Sci. Instrum. 88 (084101) (2017).

[17] B. P. MacLeod, D. K. Fork, B. Lam, C. P. Berlinguette, submitted. (2018)